\documentclass{article}
\usepackage[margin=3.2cm]{geometry}

\usepackage{amsmath}
\usepackage{amssymb}
\usepackage{mathrsfs}

\usepackage{amsthm}

\usepackage{xspace}

\usepackage{enumerate}
\usepackage{color}

\newcommand{\putawaay}[1]{}

\renewcommand{\phi}{\varphi}

\newcommand{\tuple}[1]{( #1 )}
\newcommand{\set}[1]{{\{ #1 \}}}

\usepackage{pgf}
\usepackage{pgfcore}
\usepgfmodule{plot}
\usepgfmodule{shapes}
\usepgfmodule{decorations}
\usepackage{tikz}
\usetikzlibrary{arrows,patterns,plotmarks,er,3d,automata,backgrounds,topaths,trees,petri,mindmap}

\usetikzlibrary{matrix,arrows}

\begin{document}

\title{Smart Transformations: {T}he Evolution of Choice Principles}

\author{Paolo Galeazzi $^{1}$ and Michael Franke $^{2}$\\
{\footnotesize
$^{1}$ ILLC, University of Amsterdam, The Netherlands; email: pagale87@gmail.com}\\
{\footnotesize
$^{2}$ Department of Linguistics, University of T\"{u}bingen, Germany; email: mchfranke@gmail.com}
}

\date{\today}

\maketitle

\begin{abstract}
Evolutionary game theory classically investigates which behavioral
patterns are evolutionarily successful in a single game. More recently,
a number of contributions have studied the \emph{evolution of preferences}
instead: which subjective conceptualizations of a game's payoffs give
rise to evolutionarily successful behavior in a single game
.
Here, we want to extend this existing approach even further by asking:
which general patterns of subjective conceptualizations of payoff
functions are evolutionarily successful \emph{across a class of games.
}In other words, we will look at evolutionary competition of payoff
transformations in ``meta-games'', obtained from averaging over
payoffs of single games. Focusing for a start on the class of $2\times2$
symmetric games, we show that regret minimization can outperform payoff
maximization if agents resort to a security strategy in case of radical
uncertainty.
\end{abstract}




\section{Preamble\label{sec:Preamble}}

\begin{quotation} \textit{This is what I aim at, because the point of philosophy is to start with something so simple as not to seem worth stating, and to end with something so paradoxical that no one will believe it.} \cite{Russ18} \end{quotation}

In the epistemic literature, there are two major and alternative,
sometimes competing, trends in formalizing belief: a probabilistic
approach, and a non-probabilistic approach. As the names suggest,
the probabilistic approach uses probabilities in order to model agents'
beliefs, whereas the non-probabilistic approach relies on more qualitative
structures (see for example \cite{baltsme08}).

Through this abstract we will assume that probabilistic and non-probabilistic
beliefs are just two different and compatible forms of belief. In
other words, we assume that agents may have probabilistic beliefs
in some circumstances and non-probabilistic beliefs in others. More
specifically, by probabilistic belief we mean in general that the
belief of an agent is expressed by a probability distribution, while
by non-probabilistic belief in this abstract we mean that the belief
is not representable by a single specific probability distribution.
Appealing to reasons of self-evidence and introspection, we hope this
sounds an uncontroversial assumption, ``so simple as not to seem
worth stating''. Or, at least, we hope the reader will agree that
it is much less controversial than assuming that agents have either
only probabilistic beliefs or only non-probabilistic beliefs.

\section{Intro\label{sec:Intro}}

Evolutionary game theory classically investigates which behavioral
patterns, when competing against each other, are evolutionarily stable
in a single game. More recently, a number of contributions have studied
the \emph{evolution of preferences} instead: which subjective conceptualizations
of a game's payoff function give rise to evolutionarily successful
behavior in a single game (\cite{algweib13}, \cite{DekElyYlan07},
\cite{heifshanspieg07}, \cite{OkVega01}, \cite{RobSam11}, \cite{Sam01}).
This literature is grounded on the so called \emph{indirect evolutionary
approach}. While the classical evolutionary approach is aimed at studying
whether a certain \emph{strategy} in a given game is evolutionarily
stable and robust, the indirect evolutionary approach allows to investigate
whether certain \emph{preferences }are evolutionarily successful or
not. Yet it was argued that \begin{quotation} \textit{The indirect evolutionary approach with unobservable preferences gives us an alternative description of the evolutionary process, one that is perhaps less reminiscent of biological determinism, but leads to no new results.} \cite{RobSam11} \end{quotation}

In this work we adopt a standpoint close to the one taken in evolution
of preferences, but we want to extend this existing approach even
further by asking: which general patterns of subjective conceptualizations
of payoff functions are evolutionarily successful \emph{across a class
of games. }In other words, we will look at evolutionary competition
of general payoff transformations in meta-games, obtained from averaging
over payoffs obtained from single games. Focusing for a start on the
class of $2\times2$ symmetric games, we show that regret minimization
can outperform payoff maximization if agents resort to a security
strategy in case of radical uncertainty. I.e., payoff maximization
turns out to be evolutionarily unstable under simple epistemic assumptions.

\section{Evolution of preferences\label{sec:Evolution-of-preferences}}

The standard model for studying the evolution of preferences (in particular,
we are referring to \cite{DekElyYlan07}) is built on a symmetric
two-player normal-form game $G$ with finite action set $A=\{a_{1},...,a_{n}\}$
and payoff function $\pi:A\times A\rightarrow\mathbb{R}$. This is
usually called the \emph{fitness game} since evolutionary selection
is driven by payoff function $\pi$. Players in the population represent
\emph{subjective preferences}, or \emph{subjective utility functions},
that can diverge from the \emph{objective} fitness given by $\pi$.
A subjective preference is a function $\theta_{i}:A\times A\rightarrow\mathbb{R}$,
and the set of subjective preferences is $\Theta\equiv\mathbb{R}^{A\times A}$.
Each player chooses the action that maximizes her \emph{subjective}
preference, but receives the \emph{objective} payoff defined by $\pi$.
Hence, player $i$'s action choice is determined by $\theta_{i}$,
but $i$'s evolutionary fitness is deterimined by $\pi$. 

Different authors enrich this basic picture with various features
(e.g., observability \cite{DekElyYlan07}, assortative matching \cite{algweib13},
etc.) and study the resulting effects in the dynamics. We do not argue
against any of these approaches here, but adopt a different one. Firstly,
we add a meta-game perspective by studying evolution of preferences
across a class of games. Secondly, we pay attention to the epistemic
situations of the agents and include the possibility that agents play
a security strategy in case of radical uncertainty.

\section{The model\label{sec:The-model}}

Instead of one fitness game $G$, we consider a class $\mathcal{G}$ of
fitness games. Here, we take $\mathcal{G}$ to be the class of
$2\times2$ symmetric games. We are interested in the evolutionary
competition of \emph{player types} $\tuple{\tau, e}$, conceived of as
a pair of a subjective preference type $\tau$ and an epistemic type
$e$. Let $\Lambda$ denote the set of player types.  We will enlarge on each component in turn in the following. We
take a player type to specify action choices in each $G \in \mathcal{G}$
and thus think of a player type as a \emph{choice principle}. This
allows us to study the evolutionary competition between different
subjective ways of representing a game's utilities and different ways
of using behavioral beliefs about the co-player. To keep matters
manageable, we restrict our attention to a selected subset of conceptually
relevant player types, comparing players of four different and
theoretically significant subjective preference types.

\subsubsection*{Subjective preference types}

Formally, a \emph{preference type} is a function
$\tau_{i}:\mathcal{G}\rightarrow\Theta$ from games to subjective
preferences. Let $\mathcal{T}$ be the set of preference types. We can think of preference types as transformations of
$\pi$, for any $G\in\mathcal{G}$: $\tau_{i}$ may then be understood as a player's way of
thinking across games, a red thread that relates different subjective
preferences across different games.

For perspicuity, we focus here on four conceptually
relevant transformations in $\mathcal{T}$: (i) an \emph{actual payoff
type}, whose subjective preferences coincide with actual fitness payoffs
$\pi$, (ii) an \emph{altruistic type}, whose subjective preferences
are the sum of her own fitness and that of the co-player, (iii) a
\emph{competitive type}, whose subjective preferences are her own
fitness minus that of the co-player, and (iv) a \emph{regret type},
whose subjective preferences are given by each action's \emph{regret}
(\cite{loosug82},\cite{halpass12}). Denoting the payoff function
$\pi$ of game $G$ by $\pi_{G}$, define these types as: 
\begin{itemize}
\item actual payoff type: $\forall G\in\mathcal{G},\;\tau^{\pi}(G)=\pi_{G}$;
\item altruistic type: $\forall G\in\mathcal{G},\;\forall a_{i},a_{j}\in A,\;\tau^{alt}(G)(a_{i},a_{j})=\pi_{G}(a_{i},a_{j})+\pi_{G}(a_{j},a_{i})$;
\item competitive type: $\forall G\in\mathcal{G},\;\forall a_{i},a_{j}\in A,\;\tau^{com}(G)(a_{i},a_{j})=\pi_{G}(a_{i},a_{j})-\pi_{G}(a_{j},a_{i})$;
\item regret type: $\forall G\in\mathcal{G},\;\forall a_{i},a_{j}\in A,\;\tau^{reg}(G)(a_{i},a_{j})=-(\pi_{G}(a^{\$},a_{j})-\pi_{G}(a_{i},a_{j}))$,
where $a^{\$}$ stands for the best reply to $a_{j}$ under $\pi_{G}$.%
\footnote{Formally, this is the \emph{negative} regret. We use this formulation
because it is the most convenient in this context.%
}
\end{itemize}

\subsubsection*{Epistemic types}

In full generality, an epistemic type is a general disposition to form
beliefs about the co-player's behavior. As for preference types, in this abstract we limit ourselves to a small selection of epistemic types
that are particularly interesting from a theoretical point of view. Here, we just consider two epistemic types  $e \in
\set{\bar{\mu}, \Delta(A)}$:
\begin{itemize}
\item a \emph{uniform} probabilistic belief $\bar{\mu}\in\Delta(A)$ about
  the opponent's behavior, or
\item the full set $\Delta(A)$ of all possible behavioral beliefs about
  the co-player's actions.
\end{itemize}

Hence, we are mainly focusing on two extreme epistemic
types for the moment: players can either have a probabilistic (flat) belief about the co-player's
actions, or be \emph{radically uncertain}, i.e., have no specific
probability distribution on the co-player. It would also be possible
to take into account different degrees of uncertainty, and to link
our results to the literature about ambiguity and uncertainty aversion
more tightly (\cite{GhirMar02}, \cite{GilSch89}, \cite{MacMarRus06}), but we will only consider the two extreme
cases for this abstract. 
There are many reasons why agents might be radically uncertain: lack
of cognitive capabilities, lack of information%
\footnote{Lack of information might depend for instance on the fact that players
haven't played enough rounds to learn from experience and to form
a precise probabilistic belief; alternatively, we can imagine that
players have specific probabilistic beliefs if they have already met
and know the co-player, but they do not have a single probabilistic
belief when they meet a co-player for the first time. Similarly, we
can also think that a player has specific probabilistic beliefs when
she is facing a game that she alreday played before, and imprecise
beliefs otherwise.%
}, etc. It
is to be expected that the more ``correct'' the beliefs of a player
are on average, the higher its fitness. Still, radical uncertainty may
well be considered a starting point for evolutionary selection, and so
we start our investigation there.

\subsubsection*{Choice principles}

We take player types $\tuple{\tau, e}$ to rise to \emph{choice
  principles}, i.e., systematic mappings of each game into a subset of
actions. Many possibilities
are conceivable here. To be practical, we need to, again, make
a principled selection based on theoretical relevance. For simplicity,
we assume that players apply \emph{maximin expected utility}
(\cite{GilSch89}) based on their subjective
preference type and their epistemic type.

From the perspective of \emph{Maxmin expected utility} (\cite{GilSch89}),
the behavior of an agent corresponds to maximizing the minimal expected (subjective)
utility over the \emph{set} of probability distributions that she
is holding. In our particular case, this set can either be a singleton
(a flat probability distribution $\bar{\mu}$), or the full simplex $\Delta(A)$,
i.e., the set of all possible probability distributions over the co-player's
choices (in case of radical uncertainty). In the first case, playing
an action that maximinimizes subjective expected utility is the same as maximizing subjective expected utility,
whereas playing an action that maximinimizes subjective expected utility in the second case amounts
to playing standard \emph{maximin} over the game with subjectively
transformed preferences (\cite{OsbRub94}).


Other possible construals of choice principles are conceivable, e.g.,
\emph{maximax}, the maximization of the maximal utility. Our choice of
maximin expected utility is motivated by the fact that it gives rise
to well-known decision rules. For player types $\tuple{\tau^\pi,
  \bar{\mu}}$ we have standard maximization of expected utility; for
player types $\tuple{\tau^\pi, \Delta(A)}$ we have standard maximin;
for player types $\tuple{\tau^{reg}, \Delta(A)}$ we have (positive)
regret minimization (\cite{halpass12}).

It is important to notice that player types $\tuple{\tau^\pi,
  \bar{\mu}}$ and $\tuple{\tau^{reg}, \bar{\mu}}$ are actually
behaviorally equivalent.

\medskip{}

\textbf{Remark 1.\label{Remark-1.-Probability-Maximization=00003Dminimization}
}\emph{Maximization of expected utility and minimization of expected
regret coincide: for any probabilistic belief, an action $a^{*}$
maximizes the expected utility if and only if action $a^{*}$ minimizes
the expected (positive) regret.}

\medskip{}

\noindent Nonetheless, it is important from an evolutionary point of
view to distinguish player types who conceptualize a game's
payoffs in terms of regret from those who consider the actual
payoffs $\pi$, especially when we consider evolutionary dynamics involving
mutation (see below).






\section{Results\label{sec:Results}}

In this section we present some results achievable in our set-up.
For reasons of exposition, we first focus on radical
uncertainty, and then we allow players to have both probabilistic and
non-probabilistic beliefs (i.e., to be of both epistemic types $\bar{\mu}$) and $\Delta(A)$). 

Consider a population where the eight player types introduced above ($\tuple{\tau^\pi, \bar{\mu}}$, $\tuple{\tau^{alt}, \bar{\mu}}$, $\tuple{\tau^{com}, \bar{\mu}}$, $\tuple{\tau^{reg}, \bar{\mu}}$, $\tuple{\tau^\pi, \Delta(A)}$, $\tuple{\tau^{alt}, \Delta(A)}$, $\tuple{\tau^{com}, \Delta(A)}$ and $\tuple{\tau^{reg}, \Delta(A)}$)
are present. We are interested in the question which player types will
be evolutionarily successful when repeatedly playing random symmetric
$2\times2$ games.

To address this question, we use numerical simulation to approximate
the average payoff accrued by each choice principle.
To this end, we randomly generated $50000$ symmetric $2\times2$ games by sampling i.i.d. payoffs from the natural numbers in the set $\{0,1,...,10\}$. For each sampled
game, we let all choice principles play against each other and recorded
the payoffs obtained after each play. Finally, we took the average.
Table~\ref{tab:Utils-MetaGame}
gives the resulting payoff matrix with the row type's average
payoffs against each of the column types.

\begin{table*}[ht]
\centering
\resizebox{\linewidth}{!}{%
\begin{tabular}{ccccccccc}
  \hline
 & $\tuple{\tau^{reg}, \Delta(A)}$ 
 & $\tuple{\tau^{\pi}, \Delta(A)}$ 
 & $\tuple{\tau^{alt}, \Delta(A)}$
 & $\tuple{\tau^{com}, \Delta(A)}$
 & $\tuple{\tau^{reg}, \bar{\mu}}$ 
 & $\tuple{\tau^{\pi}, \bar{\mu}}$ 
 & $\tuple{\tau^{alt}, \bar{\mu}}$
 & $\tuple{\tau^{com}, \bar{\mu}}$ \\ 
  \hline
  $\tuple{\tau^{reg}, \Delta(A)}$ & 6.629 & 6.653 & 5.806 & 7.089 & 6.636 & 6.636 & 5.793 & 7.463 \\ 
  $\tuple{\tau^{\pi}, \Delta(A)}$  & 6.455 & 6.468 & 6.067 & 6.685 & 6.462 & 6.462 & 6.065 & 6.834 \\ 
  $\tuple{\tau^{alt}, \Delta(A)}$ & 6.280 & 6.746 & 5.473 & 6.959 & 6.294 & 6.294 & 5.474 & 7.114 \\ 
  $\tuple{\tau^{com}, \Delta(A)}$ & 5.936 & 5.735 & 5.336 & 6.379 & 5.929 & 5.929 & 5.327 & 6.538 \\ 
  $\tuple{\tau^{reg}, \bar{\mu}}$ & 6.633 & 6.658 & 5.810 & 7.081 & 6.634 & 6.634 & 5.802 & 7.454 \\ 
  $\tuple{\tau^{\pi}, \bar{\mu}}$ & 6.633 & 6.658 & 5.810 & 7.081 & 6.634 & 6.634 & 5.802 & 7.454 \\ 
  $\tuple{\tau^{alt}, \bar{\mu}}$ & 6.278 & 6.750 & 5.476 & 6.953 & 6.293 & 6.293 & 5.484 & 7.112 \\ 
  $\tuple{\tau^{com}, \bar{\mu}}$ & 6.311 & 5.885 & 5.475 & 6.536 & 6.299 & 6.299 & 5.466 & 7.123 \\ 
   \hline
\end{tabular}}
\caption{Average payoff for player types in simulations of 5000
  randomly generated $2 \times 2$ symmetric games.}
\label{tab:Utils-MetaGame}
\end{table*}

\subsubsection*{Radical uncertainty}

To appreciate the following results, it helps to consider first a
restricted scenario. Assume that all players have epistemic type
$\Delta(A)$, and so play a security \emph{maximin} strategy on their
subjective representation of the game. The relevant meta-game for this
case is the top-left $4 \times 4$ payoff matrix of the full matrix in
Table~\ref{tab:Utils-MetaGame}. Essentially, we are then considering the
evolutionary competition between subjective preference types in a world of
security players.

Notice, however, that the payoffs calculated for the ``meta-game''
in Table~\ref{tab:Utils-MetaGame} depend on details of our numerical
simulation, in particular on the implicit probability with which particular
types of games are sampled. Fortunately, we can generalize the result
to an analytic statement that is independent of frequency effects,
as long as every possible game has positive occurrence probability. 

\medskip{}

\textbf{Proposition 1.\label{Proposition-1.-radical-uncertainty}
}\emph{Fix $\Lambda=\{\tuple{\tau^\pi, \Delta(A)}$, $\tuple{\tau^{alt}, \Delta(A)}$, $\tuple{\tau^{com}, \Delta(A)}$ $\tuple{\tau^{reg}, \Delta(A)}\}$,
and $\mathcal{G}$ the class of symmetric $2\times2$ games with i.i.d. payoffs sampled from the set of natural numbers $\{0,1,...,N \}$. Then $\tuple{\tau^{reg}, \Delta(A)}$
is the only evolutionarily stable type in the population.}

\medskip{}

\emph{Proof. }See Appendix.

\medskip{}

This is a conceptually noteworthy result: regret minimization
evolutionarily outperforms classic maximin on repeated plays of
$2 \times 2$ symmetric games. In other words, when playing a security strategy it is strictly better to construe
a game in terms of regrets than in terms of actual payoffs.

\subsubsection*{Full competition}

Consider next the full ``meta-game'' in
Table~\ref{tab:Utils-MetaGame}. A monomorphic population of regret
minimizers $\tuple{\tuple{\tau^{reg}, \Delta(A)}}$ is no longer
evolutionary stable; it could be invaded by expected utility maximizers of types $\tuple{\tau^{reg}, \bar{\mu}}$ and $\tuple{\tau^{\pi},
  \bar{\mu}}$. Since the latter are behaviorally equivalent, neither
is an evolutionarily stable strategy, but could at best be
\emph{neutrally stable}
\cite{Maynard-Smith1982:Evolution-and-t}. However, under our simulated
meta-game payoffs from Table~\ref{tab:Utils-MetaGame} any population
consisting entirely of $\tuple{\tau^{reg}, \bar{\mu}}$ and
$\tuple{\tau^{\pi}, \bar{\mu}}$ can be invaded by regret
minimizers. This suggests that all three types would persist under
standard evolutionary dynamics, in various relative proportions.

Simulation results of the (discrete time) \emph{replicator dynamics}
\cite{TaylorJonker1978:Evolutionary-St} indeed show that random
initial population configurations are attracted to states with only
three player types: $\tuple{\tau^{reg}, \Delta(A)}$,
$\tuple{\tau^{reg}, \bar{\mu}}$ and $\tuple{\tau^{\pi},
  \bar{\mu}}$. The relative proportions of these depend on the initial
population. This variability is eradicated if we add a small mutation
rate to the dynamics. We assume a fixed, small mutation rate
$\epsilon$ for the probability that a player's preference type
\emph{or} her epistemic type changes to another random preference type
or epistemic type. The probability that a player type randomly mutates
into a completely different player type with altogether different
preference type and epistemic type would then be $\epsilon^2$. With
these assumptions about ``local mutations'', numerical simulations of
the (discrete time) \emph{replicator mutator dynamics}
\cite{Nowak2006:Evolutionary-Dy} show that for very small
mutation rates almost all initial populations converge to a single
fixed point in which the majority of players are regret types. For
instance, with $\epsilon = 0.001$, almost all initial populations are
attracted to a final distribution with proportions:

\begin{center}
  \begin{tabular}{ccc}
    $\tuple{\tau^{reg}, \Delta(A)}$ & $\tuple{\tau^{reg},
      \bar{\mu}}$ & $\tuple{\tau^{\pi}, \bar{\mu}}$ \\ \hline
    0.279  &   0.383 &    0.281 
  \end{tabular}
\end{center}

Again, this shows that there are plausible and simple conditions under which
agents who represent the game in terms of regret may be favored by
evolutionary selection and that, more specifically, regret minimizers
are evolutionarily supported.

\subsubsection*{Variable, but correlated epistemic types}

The foregoing results were based on the implicit assumption that
players have a fixed epistemic type. Epistemic types were considered
under evolutionary pressure in tandem with preference types. In the
remainder we focus on the evolution of preference types, assuming that
players are of variable epistemic types.

For a clear-cut analytical corollary from the previous results in
Proposition 1 and Remark 1, let us assume that epistemic types are
correlated in random encounters: whenever two player types are matched
to play a game, they are always of the same epistemic type, with
positive probability that both are of either type $\bar{\mu}$ and
$\Delta(A)$.\footnote{This assumption of correlation of epistemic
  types can be motivated by the idea that some external circumstances
  of the game (or its context of presentation or occurrence)
  involuntarily trigger players into a particular epistemic type.}
Under variable, yet correlated epistemic types, it is easy to see that
preference types $\tau^{reg}$ will outperform type $\tau^{\pi}$.

\medskip{}

\textbf{Corollary 1. } \emph{Fix $\mathcal{T}=\{ \tau^{\pi}, \tau^{reg} \}$. If agents' epistemic types vary between
  encounters and both occur with positive probability, but are always
  correlated, so that co-players are always of the same epistemic type
  in any particular round of play, $\tau^{reg}$ is the only
  evolutionarily stable preference type in the population.}

\medskip{}

\emph{Proof. }See Appendix.

\medskip{}

Notice, moreover, that since Remark 1 holds for any arbitrary probabilistic belief $\mu \in \Delta(A)$, Corollary 1 also holds for any arbitrary correlated probabilistic belief $\mu$ and not only for flat belief $\bar{\mu}$. I.e., Corollary 1 holds for any possible probabilistic belief $\mu$.

\subsubsection*{Variable, uncorrelated epistemic types}

Finally, consider the case where epistemic types of players are
variable but uncorrelated. Each player has a fixed preference type, but
is of epistemic type $\Delta(A)$ with some probability
$p$. Probability $p$ is then the average probability of a player being
a security player in the population, where that does not depend on the
player's preference type. For simplicity, consider $p$ fixed for the
whole population. In that case we can compute, for each $p$, the
average payoffs of preference types in another $4 \times 4$ meta-game,
derived from the full Table~\ref{tab:Utils-MetaGame}. It turns out
that there is a very low threshold on $p$ above which regret types
dominate the evolutionary armsrace. With only a small occurrence
probability of security players $p = 0.01$, the derived meta-game
between preference types is:
\vspace{-0.2cm}
\begin{center}
  \begin{tabular}{rrrrr}
    & $\tau^{reg}$ 
    & $\tau^{\pi}$ 
    & $\tau^{alt}$
    & $\tau^{com}$ \\
    \hline
    $\tau^{reg}$ & 6.634 & 6.635 & 5.802 & 7.450 \\ 
    $\tau^{\pi}$ & 6.633 & 6.633 & 5.804 & 7.444 \\ 
    $\tau^{alt}$ & 6.293 & 6.297 & 5.484 & 7.111 \\ 
    $\tau^{com}$ & 6.295 & 6.291 & 5.465 & 7.111 \\ 
  \end{tabular}
\end{center}

Regret types are the only evolutionary stable type in this case. In
sum, even with a small probability of lacking a concrete (flat) belief
about the opponent, a subjective representation of payoffs in terms of
regret is favored by evolutionary selection.

\section{Conclusion\label{sec:Conclusion}}

The assumption that players and decision makers maximize their
preferences is central through all economic literature, and the
maximization of \emph{actual} payoffs is often justified by appealing
to evolutionary arguments and natural selection. In contrast to the
standard view, we showed the existence of player types with subjective
utilities different from the actual payoffs that can outperform types
who have subjective utilities equal to the actual payoffs.

While the literature on \emph{evolution of preference} has focused on
fixed games, or fixed types of games, we have taken a more
general approach here. We suggested that attention to ``meta-games''
is interesting, because what may be a good subjective representation
in one type of game (e.g., cooperative preferences in the Prisoner's
Dilemma class) need not be generally beneficial. In fact, it turns out
that our altruistic and competitive preference types pale in the light
of regret types. 

Taken together, we presented a variety of plausible circumstances in
which evolutionary competition between choice principles on a larger
class of games can favor subjective preference transformations
focusing on regret. 

\bibliographystyle{alpha}
\bibliography{biblio}

\end{document}